\documentclass[%
 reprint,
https://www.overleaf.com/project/5cedae5157df043b4ec4e324%showpacs,preprintnumbers,
 amsmath,amssymb,
 aps,
prl,
]{revtex4-1}

\usepackage{graphicx}% Include figure files
\usepackage{epstopdf}
\usepackage{dcolumn}% Align table columns on decimal point
\usepackage{bm}% bold math
\usepackage{lineno}
\usepackage{color}

\begin{document}

\preprint{APS/123-QED}

\title{Experimental and modelling evidence for structural crossover in supercritical CO$_2$}% Force line breaks with \\
%\thanks{A footnote to the article title}%

\author{Cillian J. Cockrell}%
\email{c.j.cockrell@qmul.ac.uk}
\affiliation{School of Physics and Astronomy, Queen Mary University of London, Mile End Road, London, E1 4NS, UK.}%\\
%\email{c.j.cockrell@qmul.ac.uk}

\author{Oliver Dicks}%
\email{o.dicks@qmul.ac.uk}
\affiliation{School of Physics and Astronomy, Queen Mary University of London, Mile End Road, London, E1 4NS, UK.}%\\
%\email{o.dicks@qmul.ac.uk}

\author{Ling Wang}%
\email{ling.wang@qmul.ac.uk}
\affiliation{School of Physics and Astronomy, Queen Mary University of London, Mile End Road, London, E1 4NS, UK.}%\\
%email{ling.wang@qmul.ac.uk}

\author{Kostya Trachenko}%
\email{k.trachenko@qmul.ac.uk}
\affiliation{School of Physics and Astronomy, Queen Mary University of London, Mile End Road, London, E1 4NS, UK.}%\\
%\email{k.trachenko@qmul.ac.uk}

\author{Alan K. Soper}%
\email{alan.soper@stfc.ac.uk}
\affiliation{ISIS Facility, STFC Rutherford Appleton Laboratory, Harwell Campus, Didcot, Oxon OX11 0QX, UK.}%\\
%\email{alan.soper@stfc.ac.uk}
 
\author{Vadim V. Brazhkin}%
\email{brazhkin@hppi.troitsk.ru}
\affiliation{Institute for High Pressure Physics, RAS, 108840, Moscow, Russia.}%\\
%\email{brazhkin@hppi.troitsk.ru}

\author{Sarantos Marinakis}%
\email{s.marinakis@uel.ac.uk}
\affiliation{School of Health, Sport and Bioscience, University of East London, Stratford Campus, Water Lane, London E15 4LZ, UK}%\\
\affiliation{Department of Chemistry and Biochemistry, School of Biological and Chemical Sciences, Queen Mary University of London,
Joseph Priestley Building, Mile End Road, London E1 4NS, UK}%\\
%\email{s.marinakis@uel.ac.uk}
%\collaboration{CLEO Collaboration}%\noaffiliation

\date{\today}% It is always \today, today,
             %  but any date may be explicitly specified

\begin{abstract}
Physics of supercritical state is understood to a much lesser degree compared to subcritical
liquids. Carbon dioxide in particular has been intensely studied, yet little is known about
the supercritical part of its phase diagram. Here, we combine neutron scattering experiments
and molecular dynamics simulations and demonstrate the structural crossover at the Frenkel
line. The crossover is seen at pressures as high as 14 times the critical pressure and is
evidenced by changes of the main features of the structure factor and pair distribution
functions.
\begin{description}
\item[Usage]
Secondary publications and information retrieval purposes.
\item[PACS numbers]
May be entered using the \verb+\pacs{#1}+ command.
% \item[Structure]
% You may use the \texttt{description} environment to structure your abstract;
% use the optional argument of the \verb+\item+ command to give the category of each item. 
\end{description}
\end{abstract}

\pacs{Valid PACS appear here}% PACS, the Physics and Astronomy
                             % Classification Scheme.
%\keywords{Suggested keywords}%Use showkeys class option if keyword
                              %display desired
\maketitle

%\tableofcontents

\section{\label{sec:intro}Introduction}

Supercritical fluids have unique properties that have led to a rich variety of applications
\cite{deben}. Rare gases, nitrogen, CO$_2$ and H$_2$O are among the most common supercritical
fluids. CO$_2$ in particular is an important greenhouse gas of Earth's atmosphere, and in its
supercritical state is the main component (97\%) in the atmosphere of Venus. Supercritical
CO$_2$ is used in a great variety of applications (see, e.g., applications in solubility,
synthesis and processing  of polymers \cite{co9,co13,co11}, dissolving and deposition in
microdevices \cite{co7}, green chemistry and solvation \cite{co6,co4,co1,co3,co10,co15,co16},
green catalysis \cite{co17,co3,co2,co51}, extraction \cite{co5}, chemical reactions
\cite{co8}, green nanosynthesis \cite{co12} and sustainable development including carbon
capture and storage \cite{co14}). It has been widely appreciated that improving fundamental
knowledge of the supercritical state is important for the reliability, scale-up and widening
of these applications (see, e.g., Refs \cite{deben,co3,co4,co51,co8,co16}).

Compared to subcritical liquids, the supercritical state is not well understood. Traditional
understanding amounted to a general assertion that this state is physically homogeneous, with
no qualitative changes taking place anywhere above the critical point \cite{deben}. The first
challenge to this view was the Widom Line (WL). Close to the critical point, the WL
characterises persisting near-critical anomalies such as the maximum in the heat capacity
\cite{stanleywidom05}, which can be used to stratify different states in the supercritical
region. A different subsequent proposal was based on the Frenkel line (FL) separating two
distinct states in the supercritical state with liquid-like and gas-like dynamics. Differently
from the WL, the FL extends to arbitrarily high pressures and temperatures (as long as
chemical bonding is unaltered), is unrelated to the critical point and exists in systems with
no boiling line or critical point \cite{pre,prl,phystoday}. The FL is also of practical
importance because it corresponds to the solubility maxima in supercritical CO$_2$
\cite{pre1}.

Here, we combine neutron scattering experiments and molecular dynamics (MD) simulations and
show evidence for the structural crossover of supercritical carbon dioxide at the Frenkel line.
The crossover extends to pressure as high as 14 times the critical pressure and is evidenced by
changes of the main features of the structure factor and pair distribution functions. The
neutron scattering experiments evidencing a crossover at highly supercritical pressure are the
first of its kind for CO$_2$.

\section{\label{sec:methods}Methods}

We recall that particle dynamics combine solid-like oscillations around quasi-equilibrium
positions and diffusive jumps between different positions below the FL, the typical character
of molecular motion in liquids \cite{frenkel}. Above the line, particle dynamics lose this
oscillatory component and become purely diffusive. This gives a practical criterion to
calculate the FL based on the disappearance of minima of the velocity autocorrelation function
(VAF) \cite{prl}. This criterion coincides with the thermodynamic criterion $c_v=2k_{\rm B}$
corresponding to the disappearance of transverse-like excitations in a monatomic system
\cite{ropp}. Since structure and dynamics are related \cite{argon}, the FL crossover was
predicted to result in a crossover of the supercritical {\it structure}.

The pressures we consider for both experiments and MD simulations are 500 and 590 bar. The FL in CO$_2$ was previously calculated using the VAF
criterion \cite{pre1}, giving us the following two state points of the predicted crossover:
(500 bar, 297 K) and (590 bar, 302 K). We recall that the FL extends to
arbitrarily high pressure and temperature above the critical point, but at low temperature it
touches the boiling line at around 0.8$T_c$, where $T_c$ is the critical temperature \cite{prl}
(note that the system does not have cohesive liquid-like states at temperatures above
approximately 0.8$T_c$ \cite{stishov}, hence crossing the boiling line at around $0.8T_c$ and
above can be viewed as a gas-gas transition \cite{prl}.) The critical point of CO$_2$ is (73.9
bar, 304.3 K), hence our state points correspond to near-critical temperatures and pressures
well above critical. In this regard, we note that the supercritical state is often defined as
the state at $P>P_c$ and $T>T_c$. This definition is loose, not least because an isotherm drawn
on ($P$, $T$) diagram above the critical point crosses the melting line, implying that the
supercritical state can be found in the solid phase. As a result, one can meaningfully speak
about near-critical part of the phase diagram only when discussing the location of the
supercritical state on the phase diagram \cite{uspehi}. As far as our state points are
concerned, they correspond to temperatures much higher than the melting temperature and
pressures extending to 14 times the critical pressure where near-critical anomalies are
non-existent \cite{uspehi}.

%\subsection{\label{sec:expt}Experimental}

A cylinder of carbon dioxide was obtained from BOC, CP grade, and used without further
purification. The pressure of the cylinder was around 50 bar and a SITEC intensifier and a
SITEC hand pump gas was used to raise the pressure. Capillaries were used to connect
intensifier manifold system to the cell. The flat plate pressure cell was made from an alloy of
Ti and Zr in the mole ratio 0.676:0.324, which contributes almost zero coherent scattering to
the diffraction pattern \cite{soper17}. The cell consisted of a flat section that was 12\,mm
thick and had four 6\,mm diameter holes running through it, so the occupied gas space was 6\,mm
thick and the wall thickness was 3\,mm either side. The container was placed at right angle to
the neutron beam, which was approximately 30 mm x 30 mm in cross section. A bottom loading closed cycle helium
refrigerator was used to control the temperature within $\pm$ 1 K, using He exchange gas at
$\sim$20 mbar to provide temperature uniformity. The employed temperatures and pressures are
shown in Table~\ref{tbl:ptx}, where the densities were calculated from the data available in
the NIST database \cite{nist}.

\begin{table}[h]
\small
\caption{\ $T-P-d$ state points for neutron scattering measurements. The values of $d$ are taken from \cite{nist}. }
\label{tbl:ptx}
\begin{tabular*}{0.48\textwidth}{@{\extracolsep{\fill}}ccccc}
\hline
$T_{\rm exp}$ (K) & $P_{\rm exp}$ (bar) & $d$ (g/mL) & $P_{\rm exp}$ (bar) & $d$ (g/mL) \\
    \hline
    250           & 500                 & 1.1676     & 590                 & 1.1821\\
    270           & 500                 & 1.1131     & 590                 & 1.1306\\
    290           & 500                 & 1.0573     & 590                 & 1.0784\\
    310           & 500                 & 1.0003     & 590                 & 1.0257\\
    330           & 500                 & 0.9426     & 590                 & 0.9729\\
    340           & 500                 & 0.9137     & -                   & -     \\
    350           & 500                 & 0.8848     & 590                 & 0.9204\\
    360           & 500                 & 0.8560     & -                   & - \\
    370           & 500                 & 0.8276     & 590                 & 0.8688\\
    380           & 500                 & 0.7996     & 590                 & 0.8436\\
    390           & 500                 & 0.7722     & 590                 & 0.8188\\
\hline
\end{tabular*}
\end{table}

\begin{figure}[!h]
\includegraphics[width=8.0cm,angle=0.]{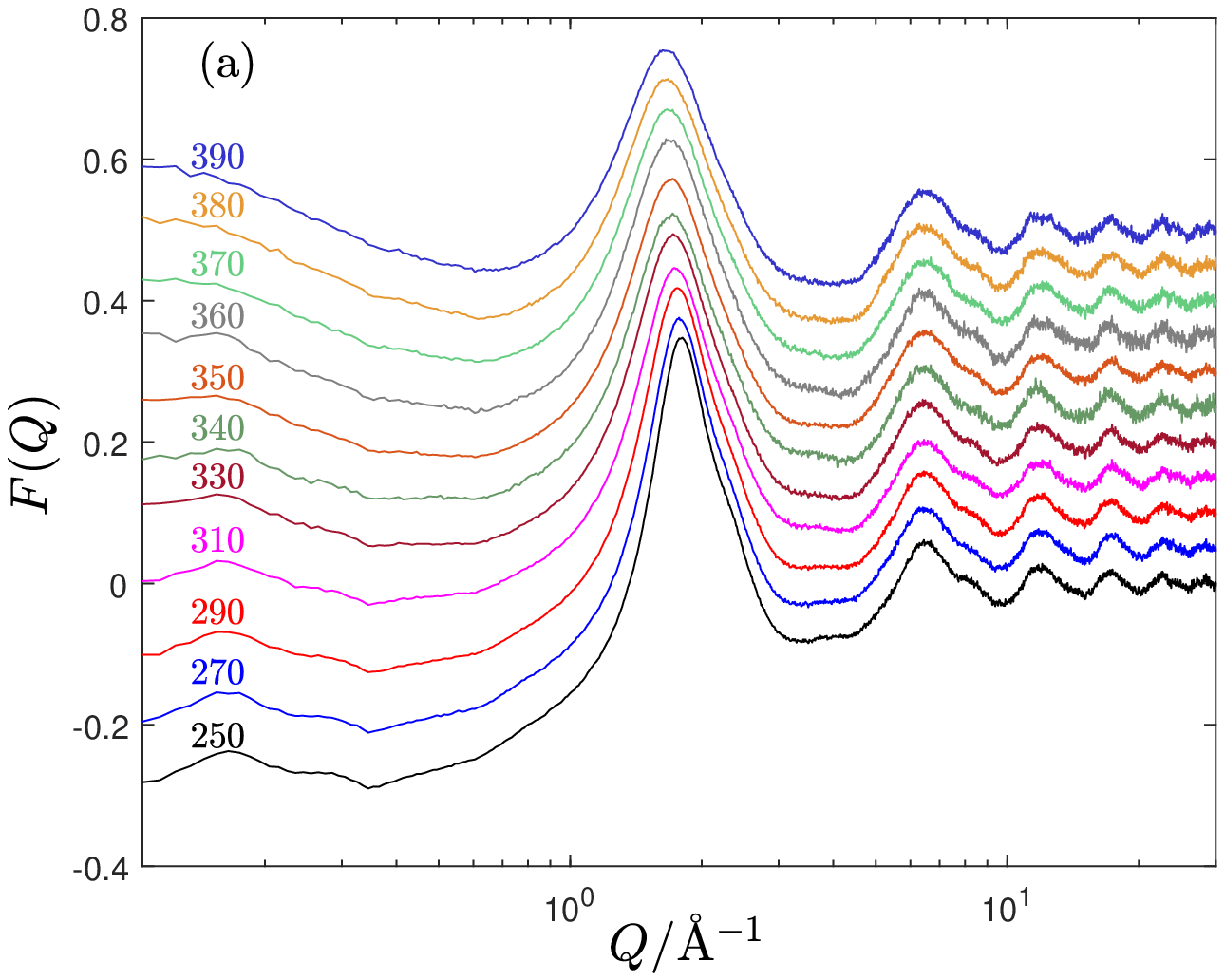}
\includegraphics[width=8.0cm,angle=0.]{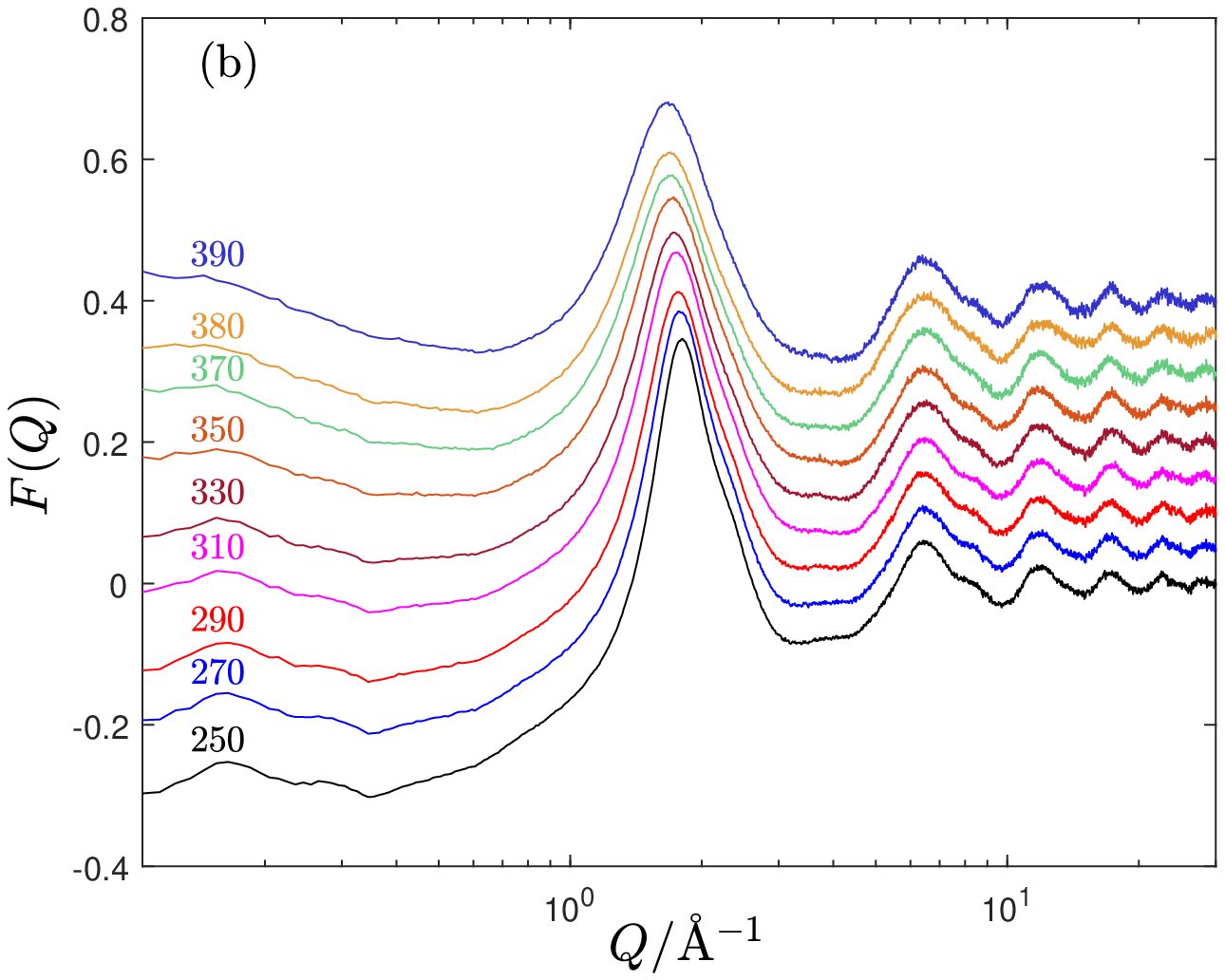}
\includegraphics[width=8.0cm,angle=0.]{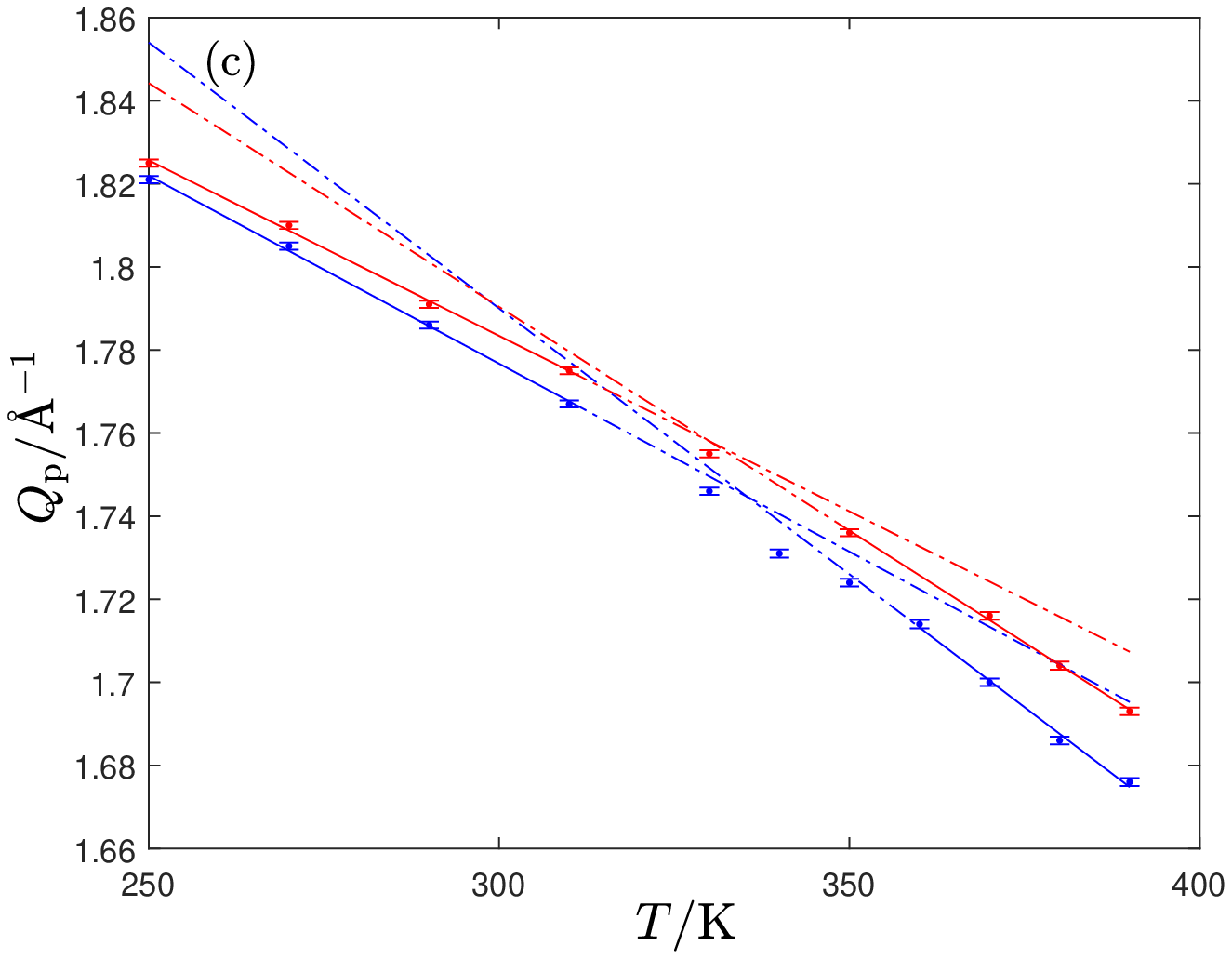} 
\caption{Weighted sum of the experimental weighted partial structure factors $F(Q)$ for CO$_2$
at (a) 500\,bar and (b) 590\,bar. The curves for higher temperatures have been
shifted along the y-axis by 0.05 per set. (c) Position of the maximum of the first peak in
the total (weighted sum) experimental structure factor as a function of
temperature for CO$_2$ at 500\,bar (blue) and 590\,bar (red). The straight lines are visual
guides.}
\label{fig:tsf}
\end{figure}

Total neutron scattering measurements were performed on the NIMROD diffractometer at the ISIS
pulsed neutron source \cite{bowron10}. Absolute values of the differential cross sections were
obtained from the raw scattering data by normalising the data to the scattering from a slab
of vanadium of known thickness, and were further corrected for background and multiple
scattering, container scattering and self-attenuation, using the Gudrun data analysis program
\cite{gudrun}. Finally the data were put on absolute scale of barns per atom per sr by dividing
by the number of atoms in the neutron beam (1 barn = $10^{-28} \rm{m}^2$). 

\begin{eqnarray}
F(Q) = \frac{1}{9}b_{\rm C}^{2}H_{\rm CC}(Q)+\frac{4}{9}b_{\rm O}^{2}H_{\rm OO}(Q)+\frac{4}{9}b_{\rm C}b_{\rm O}H_{\rm CO}(Q) \nonumber \\ 
\label{eqn:fq}
\end{eqnarray}
where $b_{\alpha}$ is the neutron scattering length of atom $\alpha$, and the partial structure
factor $H_{\alpha\beta}(Q)$ is the three-dimensional Fourier transform of the corresponding
site-site radial distribution function:

\begin{eqnarray}
H_{\alpha\beta}(Q) = 4\pi\rho\int_{0}^{\infty}r^2(g_{\alpha\beta}(r)-1)\frac{\sin Qr}{Qr} dr
\label{eqn:hq}
\end{eqnarray}
and $\rho$ is the atomic number density. Note that the $H_{\rm OO}(Q)$ and $H_{\rm CO}(Q)$
terms include both the intra- and inter-molecular scattering. The results are shown in
Fig.~\ref{fig:tsf}.

% 1

%\subsection{\label{sec:comput}Computational}

The molecular dynamics (MD) simulation package DL\_POLY \cite{dlpoly06} was used to simulate a
system of 30752 CO$_2$ particles with periodic boundary conditions. The potential for CO$_2$
is a rigid-body non-polarizable potential based on a quantum chemistry calculation, with the
partial charges derived using the distributed multipole analysis method \cite{gao17}. The
electrostatic interactions were evaluated using the smooth particle mesh Ewald method in MD
simulations. The potential was derived and tuned using a large suite of energies from {\it ab
initio} density functional theory calculations of different molecular clusters and validated
against various sets of experimental data including phonon dispersion curves and $PVT$ data.
These data included solid, liquid and gas states, gas-liquid coexistence lines and extended to
high-pressure and high-temperature conditions \cite{gao17}. We also used another
rigid-body non-polarizable potential developed by Zhang and Duan \cite{rigidpotential} and
found the same results.

The MD systems were first equilibrated in the constant pressure and temperature ensemble for
500 ps. The data were subsequently collected from production runs in the constant energy and
volume ensemble. In order to reduce noise and see the crossover clearly, data were averaged
over 500,000 frames, involving production runs of further 500 ps.

\section{\label{sec:results}Results and discussion}

%\subsection{\label{sec:intestru}Intermolecular structure}

Before analyzing the data, we recall that the FL corresponds to the qualitative change of
particle dynamics: from combined solid-like oscillatory and diffusive dynamics below the line
to purely diffusive gas-like dynamics above the line. Therefore, the supercritical structure
is predicted to show the crossover between the liquid-like and gas-like structural
correlations. This is predicted to be the case for functions characterizing the structure,
such as pair distribution function (PDF) and structure factor (SF). In our analysis, we focus
on meaningful features such as maxima positions of PDFs and SFs. 

The experimental weighted sum of the partial structure factors are plotted in Fig.
\ref{fig:tsf} for two pressures. We plot the first peaks position of SFs vs temperature in Fig.
\ref{fig:tsf} and observe that it undergoes the crossover at temperatures close to 320 K and
around 12\% larger than the FL crossover temperature predicted from the VAF criterion 
mentioned earlier.

\begin{figure}[!h]
\includegraphics[width=8.0cm,angle=0.]{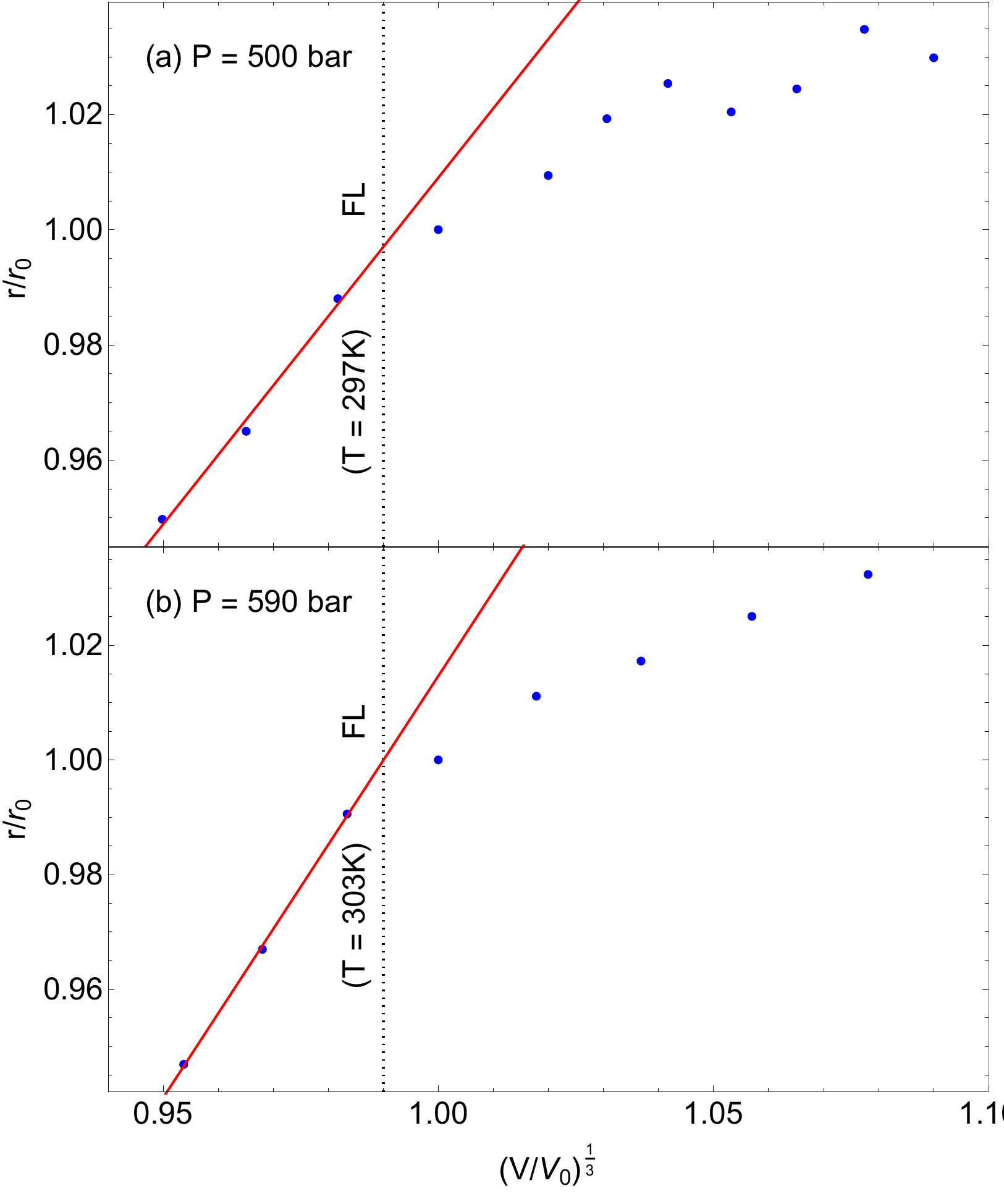}
\caption{First peak positions of experimental weighted sum PDF for 
CO$_2$ at (a) 500 bar and (b) 590 bar as a function of volume. The vertical dashed lines 
show the reduced volumes at the FL, and the straight lines are visual guides.}
\label{sqpp}
\end{figure}

\begin{figure}[!h]
\includegraphics[width=8.0cm,angle=0.]{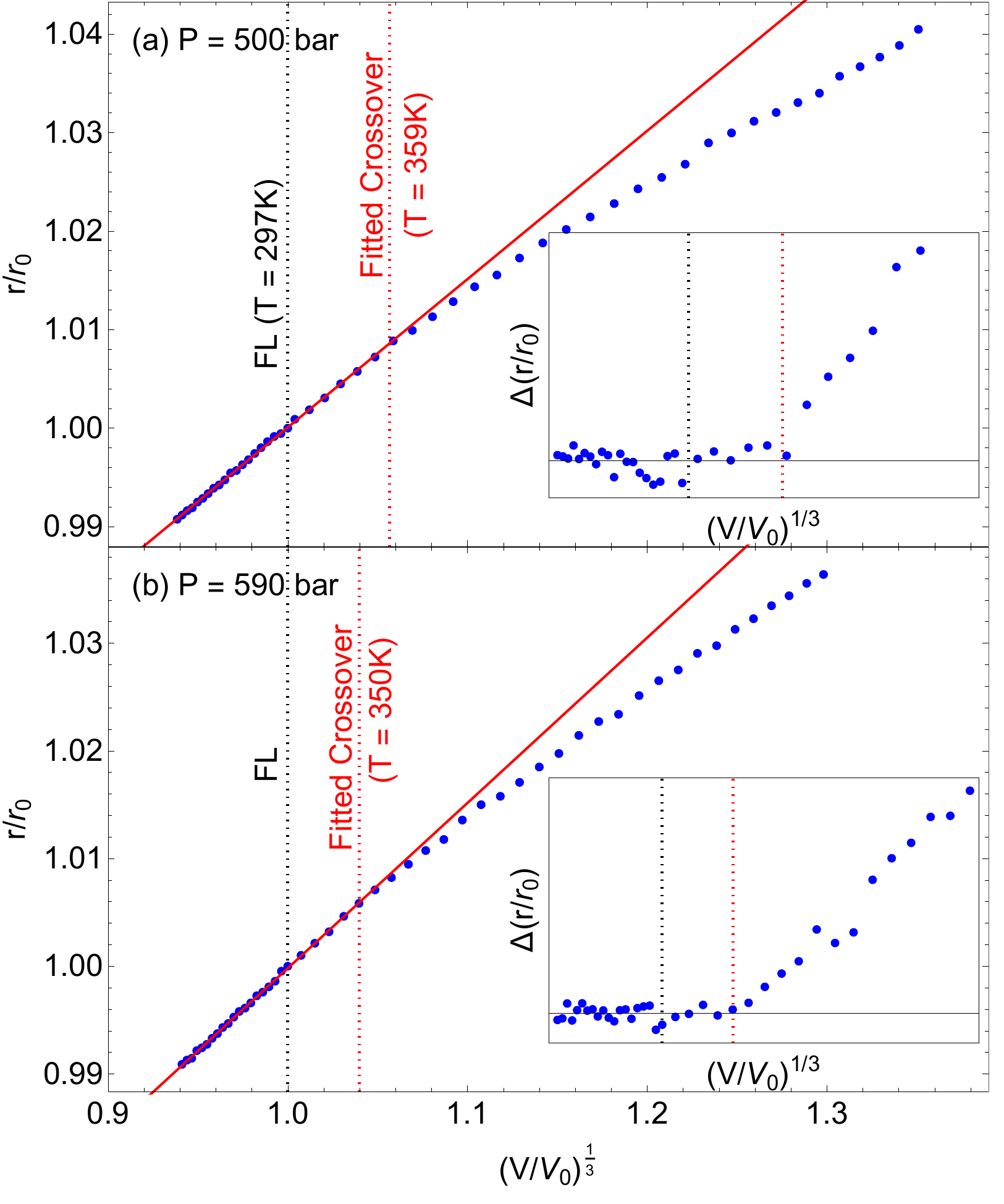}
\caption{First peak position of simulated C-C PDF. The straight lines are fitted to data 
below the FL and serve as visual guides. The vertical dashed lines show the fitted crossover volume 
and the volume at the FL. The insets show the relative trend of the residuals of the linear fit.}
\label{rdfpeakpos}
\end{figure}

The SFs were Fourier transformed to obtain the experimental PDF. As with previous experimental and modelling results on Ar \cite{argon}, 
Ne \cite{ne}, CH$_4$ \cite{methane}, and especially on H$_2$O \cite{water} 
pronounced changes of first peak position in the PDF with temperature are observed, indicating a well-defined crossover.  When a system
is compressed or expanded, one expects the first nearest-neighbour distance, $r_\mathrm{fnn}$ (given by the radial position of the first peak in g(r)), and the system's
``length" ($V^{1/3}$) to be proportional to each other unless the system undergoes a structural change. 
In other words the system structure undergoes uniform compression. The first PDF peak position divided by the position 
($r_0$) at the Frenkel temperature vs the cube root of the volume  
divided by the volume ($V_0$) at the same reference temperature is shown in Fig. \ref{sqpp}. For a system undergoing uniform compression, $V/V_0$ and $r/r_0$ will be equal. If there is a phase transition at higher densities, as there is in liquids across the melting line, this rule cannot be extrapolated down to arbitrarily low volumes and hence there will be an intercept: $V^\frac{1}{3} = \alpha r + \beta$, upon which the gradient of $V/V_0$ vs. $r/r_0$ will depend. However as long as no structural changes occur, the gradient will remain constant. Specifically, in a simple cubic crystalline solid (atomic packing fraction 0.52) the constant of proportionality between $V/N$ and $r$ is unity, in a FCC lattice (packing fraction 0.74) the constant is 0.89, and in a diamond cubic lattice (packing fraction 0.34), the constant is 1.2. In gases, the fnn 
distance is largely determined by the size, geometry, and interaction of the constituent molecules 
(see, e.g., \cite{ziman}) rather than the density. This linear relationship has been experimentally
observed in molten group 1 elements \cite{katayama2, katayama3} and liquid CS$_2$ \cite{katayama1}.
The fnn distance is most readily extracted from the partial C-C PDFs. The experimental data give
the total PDF, but the peak corresponding to the fnn distance is not profoundly changed, therefore
the total PDF gives a qualitative approximation of the fnn distance.
In Fig. \ref{sqpp} we observe the crossover of the first PDF peak at a temperature within 10\% of
the predicted crossover temperature at the FL, signified by the change of gradients. This qualitative behaviour is seen more clearly in the MD results (Fig. \ref{rdfpeakpos}).

% 2

% 3
\begin{figure}[!h]
\includegraphics[width=8.0cm,angle=0.]{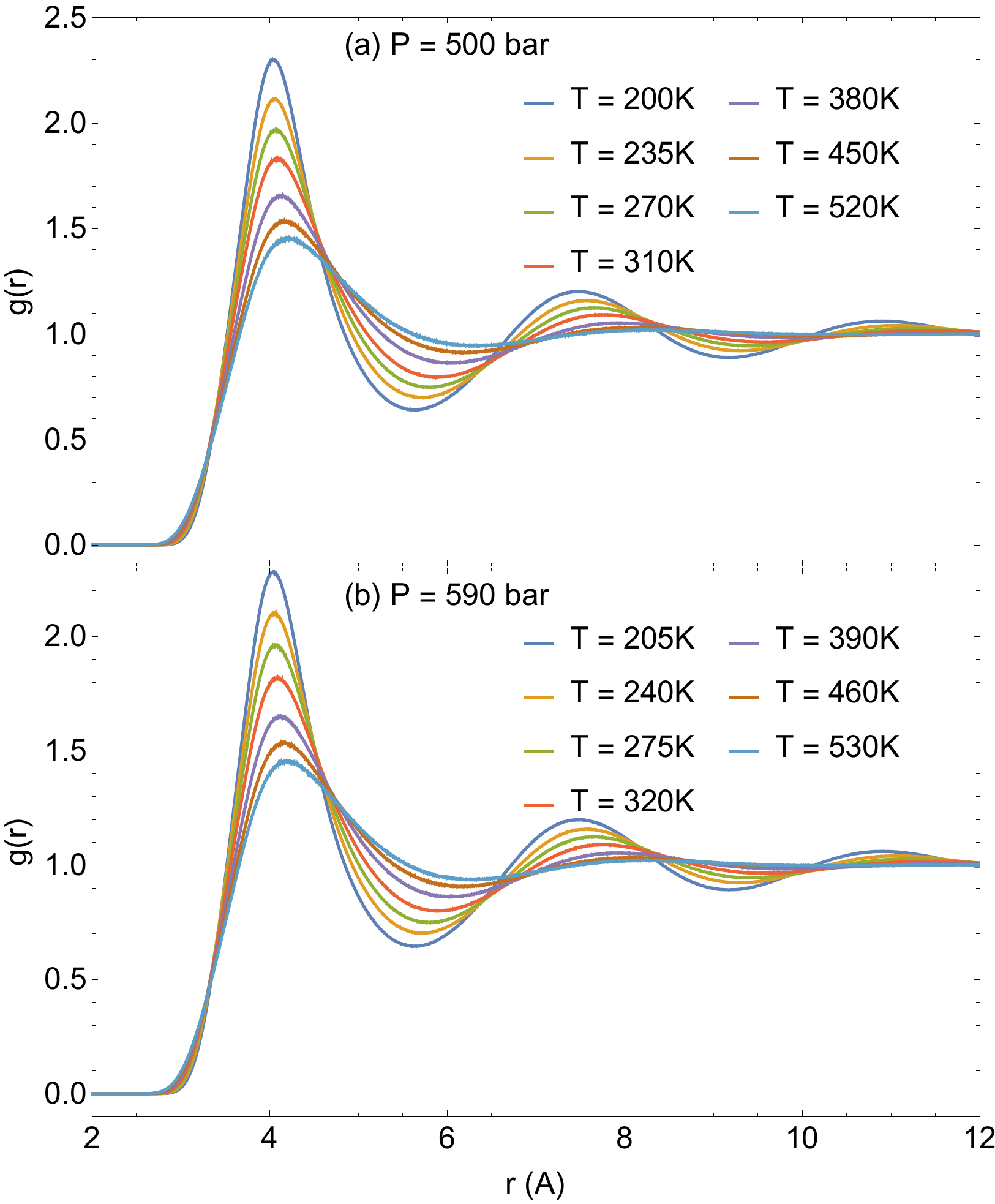}
\caption{(Colour online) Evolution of the simulated C-C pair distribution functions with
temperature at (a) 500 bar and (b) 590 bar.}
\label{tsf}
\end{figure}

We now discuss the MD results. Examples of C-C PDFs from MD simulations are shown in Fig. \ref{tsf}. We observe a reduction in height, and corresponding broadening of peaks with increasing temperature as expected. The steepness of the first peak is related with the softness of the effective intermolecular potential, and its reduction can be quantitatively related to the reduction of the viscosity \cite{shearline}. Fig. \ref{rdfpeakpos} displays the radial positions of the first PDF  peaks as a function of volume, as discussed
above, which shows a crossover at densities near the FL. Because of the reduced noise and abundance of temperature
points we can perform statistical analysis of the data to quantify the crossover. We see the same behaviour, including
a much clearer crossover, for both pressures. The constant of proportionality between $V/N$ and $r$ is $\approx 3.4$, implying a much more open arrangement than the crystal systems quoted above. This is in accordance with the density of CO$_2$ at the FL (23346 mol/m$^3$), less than half that of water (56501 mol/m$^3$) at the same pressure and temperature. In order to quantify the crossover, we fitted the data to two different
types of function. The first was a single functional dependence over the entire range. In order to avoid the
extrapolation errors associated with high order polynomials, the trial functions we used were quadratic, or log plus
linear: $f(x) = a + b x + c \log(x)$, with $a$, $b$, and $c$ the fitting parameters. The second set of functions 
was linear below a certain crossover volume $V_c$, and either quadratic or log plus linear above that volume 
(\textit{i.e.} a piecewise function): $f(x) = \Theta(V_c-V)[a + b x] + \Theta(V-V_c)[\alpha + \beta x + \gamma \log(x)]$,
with $\Theta[V]$ the Heaviside step function and $a$, $b$, $\beta$, $\gamma$, and $V_c$ the fitting parameters ($\alpha$ 
depends on the other parameters in order to ensure continuity of the function). 

Generally speaking, adding more parameters
to a fitting function improves the numerical quality of the fit. \textit{A priori}, one can penalise having too many
parameters - this prevents the extreme situation of a perfect fit acquired using a piecewise function with a number of
subdomains equal to the number of data. The two closely related quantitative measures of goodness of fit with penalty
terms for the number of parameters are the AIC (Akaike Information Criterion) \cite{AIC} and BIC (Bayesian Information
Criterion) \cite{BIC}. Applied to our data, at both pressures and with the quadratic and log plus linear variants, the
AIC and BIC were substantially lower than $-10$ below those for the single function, representing a decisive
preference for two different functional dependences above and below a certain volume ($V_c$). This volume is shown in
the vertical dotted line (Fig. \ref{rdfpeakpos}) and corresponds at both pressures to a temperature close to 350 K,
which is within 12-15\% of the predicted crossover value. Also plotted as insets in Fig. \ref{rdfpeakpos} are the
residuals of the low-volume linear fits which show a sharp and sudden increase above the crossover volume, which would
not be the case if we had simply interpolated a straight line between non-linear data.

\begin{figure}
\includegraphics[width=8.0cm,angle=0.]{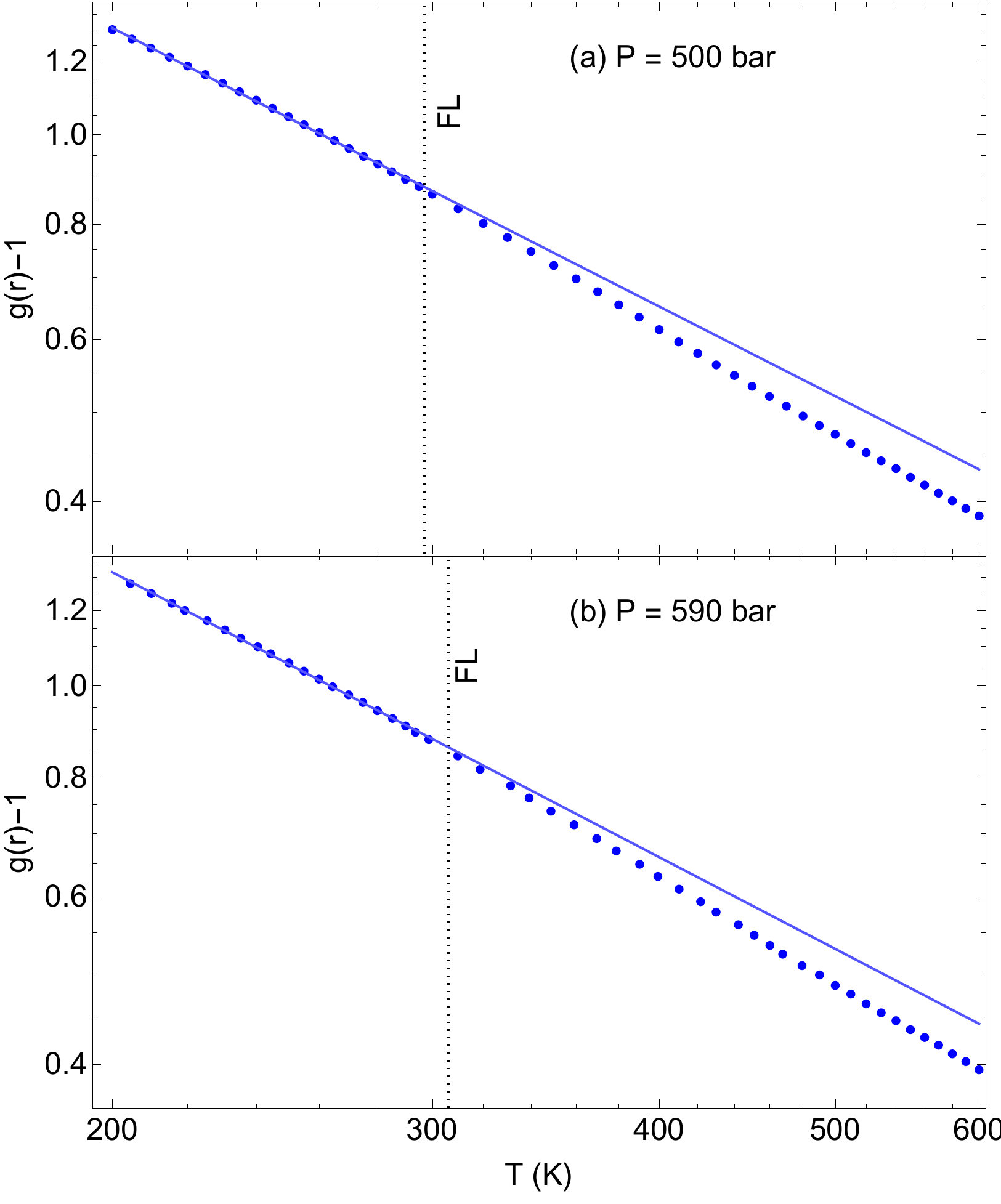}
\caption{First peak height of C-C PDF for simulated CO$_2$ at (a) 500 bar and (b) 590 bar as a 
function of volume. The vertical dashed lines show the temperatures of the predicted crossover, 
and the solid lines show the fitted straight lines.}
\label{rdfpeakheights}
\end{figure}

Fig. \ref{rdfpeakheights} shows theoretical PDF peak heights. We note that the PDF peak heights of a solid $h = g(r_\mathrm{peak}) - 1$ are
predicted \cite{tld, frenkel} to have a power-law relationship with temperature, resulting in the following relation:
$\log h \propto - \log T$ with $h = g(r)-1$ at the peak. The same relation can be argued to apply to liquids
below the FL where the solid-like oscillatory component of molecular motion is present \cite{argon}. This is because for small displacements the energy is roughly quadratic and the displacement distribution will be Gaussian. The height of a Gaussian distribution follows a power-law relationship with its variance, and thus with temperature. The peak heights in Fig. \ref{rdfpeakheights} clearly show the crossover at the FL, with the observed
crossover temperatures differing from the predicted ones by about 7-15\%. This is in agreement with the width of the FL
crossover seen experimentally and modelling on the basis of structural and thermodynamic properties \cite{ne,cv}.

% 4

% 5

%\section{\label{sec:conclu}Conclusions}

Before concluding, we note that previous experiments detecting the structural crossover at the FL involved X-ray
scattering in supercritical Ne \cite{ne}, the combination of X-ray with Raman scattering in supercritical CH$_4$
\cite{methane}, and the combination of neutron and Raman scattering in supercritical N$_2$ \cite{n2}
. Only one small-angle neutron scattering experiment had been used to study the FL in CO$_2$ in the
vicinity of the critical point only \cite{droplet}. Our current neutron scattering experiment detecting the crossover at the
FL at highly supercritical pressures is the first of its kind and importantly widens the range of techniques used to
detect the FL. It will stimulate further neutron scattering experiments in important systems such as 
supercritical H$_2$O where a pronounced crossover at the FL was recently predicted on the basis of MD simulations
\cite{water}.

In summary, our combined neutron scattering and molecular dynamics simulations study has
detected the structural crossover in CO$_2$ at pressures well above the critical pressure and
temperatures well in excess of melting temperature. The crossover is seen in the main features
of the SF and PDFs and corresponds to the predicted crossover at the FL. 
Apart from the fundamental importance of understanding the supercritical state, 
the FL corresponds to the solubility maxima of several solutes in supercritical CO$_2$
\cite{pre1} and is therefore of practical importance.

%\section*{Conflicts of interest}
%There are no conflicts to declare.

%\section*{Acknowledgements}
Neutron beam time at ISIS and project funding was provided by the Science and Technology
Facilities Council (RB1720056). We acknowledge Chris Goodway, Thomas Headen and Damian 
Fornalski (Rutherford Appleton Laboratory) for help with the experiments. This research
utilized Queen Mary's MidPlus computational facilities, supported by QMUL Research-IT,
http://doi.org/10.5281/zenodo.438045.

%\bibliography{rsc.bib} 
%\bibliographystyle{rsc} 
%\bibliographystyle{apsrev4-1}

\end{document}